# Thermo-mechanic-electrical coupling in phospholipid monolayers near the critical point

D. Steppich[a], J. Griesbauer[a,b]*,
T. Frommelt[a], W. Appelt[a], A. Wixforth[a], M. F. Schneider [b]**

[a]University of Augsburg, Experimental Physics I, Universitätstsr. 1, D-86159 Augsburg, Germany

[b]Boston University, Mechanical Engineering, 110 Cummington Street, Boston, MA, 02215 (USA)

* Corresponding *author for electrical measurements and calculations:*

*Josef Griesbauer*
*University of Augsburg*
*Experimentalphysik 1*
*Universitätsstr. 1*
*86159 Augsburg (Germany)*
*Josef.Griesbauer@physik.uni-augsburg.de*

*** Corresponding author:*

*Dr. Matthias F. Schneider*
*Boston University*
*Mechanical Engineering,*
*110 Cummington Street,*
*Boston, MA, 02215 (USA)*
*mfs@bu.edu*

**Abstract**

Lipid monolayers have been shown to represent a powerful tool in studying mechanical and thermodynamic properties of lipid membranes as well as their interaction with proteins. Using Einstein's theory of fluctuations we here demonstrate, that an experimentally derived linear relationship both between transition entropy $S$ and area $A$ as well as between transition entropy and charge $q$ implies a linear relationships between compressibility $\kappa_T$, heat capacity $c_\pi$, thermal expansion coefficient $\alpha_T$ and electric capacity $C_T$. We demonstrate that these couplings have strong predictive power as they allow calculating electrical and thermal properties from mechanical measurements. The precision of the prediction increases as the critical point $T_C$ is approached.

**Introduction**

Lipid monolayers have been of major interest for a variety of reasons. They contain many interesting aspects of physics in two dimensions, allowing us to study intermolecular interactions between lipids and/or proteins and mimic one leaflet of a bio membrane in which proteins can be incorporated fairly easy. Moreover, in contrast to most thin film physics experiments, the majority of measurements can be performed without a vast amount of technical equipment and preparation [1]. The prepared monomolecular film can then be transferred on a solid support for further investigation [2]. Most macroscopic conditions (pressure, temperature, concentration etc.) are easily accessible and thermodynamic properties can be measured straight forwardly. Furthermore, using the appropriate thermodynamic relationships, the measured quantities can be related to the heat of transition and several response functions like compressibility, thermal expansion coefficient, electrical capacity and, as will be shown in this paper, even heat capacity.

As was demonstrated earlier for lipid vesicles [3], there exists a linear relationship between the excess isothermal compressibility $\Delta\kappa_T$ and the heat capacity $\Delta c_p$ at constant pressure,

$$\Delta\kappa_T \propto \Delta c_P \qquad (0)$$

This relationship implies a pronounced maximum of the compressibility in the phase transition regime, following the well known profile of $\Delta c_p$. However, there is no physical explanation of the origin of this relationship to date.

Here, we show that the experimentally established proportionality between entropy and area and entropy and charge in combination with Einstein's theory of fluctuations, consequently leads to linear relationships between compressibility $\kappa_T$, heat capacity $c_\pi$, thermal expansion coefficient $\alpha_T$ and electric capacity $C_T$. This allows to calculate the corresponding heat capacity of lipid monolayers which is shown to be in good agreement to calorimetric data found for lipid vesicle suspensions [3].

**Material and Methods**

The Phospholipids DMPC (1,2-Dimyristoyl-*sn*-Glycero-3-Phosphocholine), DPPC (1,2-Dipalmitoyl-*sn*-Glycero-3-Phosphocholine), DMPG (1,2-Dimyristoyl-sn-Glycero-3-[Phospho-rac-(1-glycerol)]) and DOTAP (1,2-Dioleoyl-3-Trimethylammonium-Propane) were purchased from Avanti Polar Lipids (Al, USA). Lipids were dissolved in pure chloroform and used without further purification. For measuring pressure-area- and potential-area-isotherms we used film balances of Nima Technologies (Coventry, England), which were partly equipped with a Kelvin Probe Surface Potential of TREK (New York, USA) for measuring the surface potential of the lipid films. 500ml of double distilled water is filled into a trough and the lipid-chloroform solution is spread onto the water surface. After evaporation of the solvent the film was compressed with a speed of approximately 3cm$^2$/min or when accounting for the through size 0.4*10$^{-16}$/min per lipid. At the same time the lateral pressure of the surface is measured using a Wilhelmy plate and the surface potential is recorded if desired. To calculate the compressibility we subdivided the pressure-area curve in three sections and fitted each section by a polynomial function of 7$^{th}$ degree or higher to assure only minimal deviations (less than the experimental error bar) from the experimental curve. The fits were then merged together to reach continuously differentiable plots. For conserving the stronger fluctuations of the surface potential curves, numerical derivations were made for calculation of the electrical capacity.

**Results and Discussion**

**Experimental Results**

*Linear relationship between area, entropy and charge*

In Figure 1 a) and b) a set of pressure-area isotherms is shown along with the calculated isothermal compressibility [4],

$$\kappa_T = -\frac{1}{A}\frac{\partial A}{\partial \pi}\bigg|_T \qquad (1).$$

Here, $A$ is the area per molecule and $\pi$ the lateral pressure, which might not be mistaken with the global pressure $p$ as denoted in $c_p$. Additionally a set of surface potential-area isotherms ($\Psi - A$) are plotted in Figure 1 c) to demonstrate their similarity with the typical pressure-area isotherms (figure 1a) especially in the phase transition regime, whereas this is a well known behaviour [5, 6, 7].

From Figure 1 a) and b), the change in phase transition pressure $\pi_t$ as a function of temperature $T$ can be calculated and is plotted in Figure 2a. A linear relationship $\pi_t(T)$ as has been reported before [8,9] can be resolved. Interestingly the same behaviour is resembled for the transition potentials $\Psi_t(T)$ shown in Figure 2b indicating a more general character of the observed linearity, with a coupling of the thermodynamic observables studied.

When examining the pressure-area data over a wider temperature range, one finds that this linear relationship is more pronounced in the high temperature regime as the critical point is approached, but departs clearly for lower temperature (Fig. 2a). This has been found earlier, e.g. by Albrecht et al [8] and Hato et al. [10]. Qualitatively, the observed linearity seems to be independent of the nature of the lipid head group and could be reproduced for different lipids (DMPC, DMPG, DOTAP) and lipid mixtures of three different lipids (Fig. 2a). The linear character is further supported by recent results [11] where the same qualitative relationship was found for a whole set of synthesized glycolipids containing different size sugar (lactose) headgroups and no phosphoric acid.

**Theoretical Results**

The last section has experimentally shown that the change in phase transition pressure $\pi_t$ and potential $\Psi_t$ with temperature $T$ occurs linear in the regime of the critical point. As a consequence, the change in area, charge and entropy during the phase transition are linearly related according to the Clausius-Clayperon equation [5]:

$$\left.\frac{\partial \pi_t}{\partial T}\right|_{Trans} = \frac{\Delta S}{\Delta A} = const. = B_\pi \qquad (2a),$$

$$\left.\frac{\partial \Psi_t}{\partial T}\right|_{Trans} = \frac{\Delta S}{\Delta q} = const. = B_\Psi \qquad (2b)$$

and therefore

$$\Delta S = B_\pi \Delta A \qquad (3a),$$

$$\Delta S = B_\Psi \Delta q \qquad (3b).$$

For the transition pressures this relationship seems to be well conserved and becomes more accurate in the vicinity of the critical point.
When approaching the critical point the double minimum of the free energy gradually disappears and changes into a single, rather flat minimum. In the near vicinity the potential can therefore be approximated by $S = S_0 + \sum \frac{\partial S}{\partial x_i} x_i + \sum \frac{\partial^2 S}{\partial x_i \partial x_j} x_i x_j$. The corresponding fluctuations are:

$$\langle x_i x_j \rangle = -k_B \left( \frac{\partial^2 S}{\partial x_i x_j} \right)^{-1}. \qquad (4)$$

Here, $x_i$, $x_j$ are extensive thermodynamic variables (area, charge, particle number etc.), $k_B$ is the Boltzmann constant and $S(x_i, x_j)$ is the proper thermodynamic (entropy) potential. We have recently shown [12, 13], that $S$ of the monolayer can indeed be considered as decoupled from the bulk, which holds for adiabatic properties as well as for reversible fluctuations down to the nanoscale. Now using equation 4, one gets:

$$\langle \delta A^2 \rangle = A k_B T \kappa_T \qquad (5a),$$

$$\langle \delta q^2 \rangle = k_B T \left.\frac{\partial q}{\partial \Psi}\right|_{T,A} = k_B T C_T \qquad (5b),$$

$$\langle \delta S^2 \rangle = k_B c_\pi \qquad (5c),$$

where $C_T$ is the electrical capacity. For the fluctuation correlations between area and entropy

$$\langle \delta S \delta A \rangle = A k_B T \alpha_\pi. \tag{6a}$$

$$\langle \delta S \delta q \rangle = k_B T \left. \frac{\partial q}{\partial T} \right|_{\Psi,A} \tag{6b}$$

In other words, area fluctuations are proportional to the isothermal compressibility $\kappa_T$, entropy fluctuations are proportional to the isobaric heat capacity $c_\pi$ and charge fluctuations are proportional to the electrical capacity $C_T$. The correlated fluctuations are proportional to the isobaric area expansion coefficient $\alpha_\pi$ or the derivation of the charge $\left. \frac{\partial q}{\partial T} \right|_{\Psi,A}$. We like to note here that while equ. 5 and 6 are accepted textbook representations, the physically interpretation of a fluctuating entropy potential (equ. 6a and b) is physically not clear to these authors.

Under the assumption that the linear relationship in equation 3 holds also for small fluctuations ($\Delta \sim \delta$) one finds using equ. 3, equ. 5 and 6:

$$\Delta c_\pi = B_\pi^2 A T \Delta \kappa_T \tag{7}$$
$$\Delta c_\pi = B_\pi A T \Delta \alpha_\pi \tag{8}$$
$$\Delta c_\pi = B_\Psi^2 T \Delta C_T \tag{9}$$

and therefore

$$\Delta \kappa_T \propto \Delta \alpha_\pi \propto \Delta c_\pi \propto \Delta C_T. \tag{10}$$

Partly, for thermal and mechanical properties, this was already put forward by Heimburg (3) for lipid vesicle suspensions. He demonstrated that one can assume $\Delta A \sim \delta A$ when the 'excess' heat capacity $\Delta c_\pi$ and compressibility $\Delta \kappa_T$ are proportional for a wide $T$-range, the latter of which has been derived experimentally.

Eq. 10 states, that the 'excess' response functions exhibit a simple linear relationship, which predicts, according to Fig. 1, a pronounced maximum in area expansion coefficient, the electrical capacity and the heat capacity close to the phase transition regime.

In the following section we will compare our theoretical predictions with our experimental results.

**Discussion**

*Coupling between thermal and elastic properties during phase transition*

In figure 3a the area per molecule as a function of $T$ is plotted for 3 different $\pi$-values. When calculating the corresponding $\alpha_\pi$ using $\alpha_\pi = \frac{1}{A} \left. \frac{\partial A}{\partial T} \right|_\pi$ (fig. 3b) a pronounced maximum with a shoulder on the low temperature site and a steep decrease in the liquid expanded phase is found. This is very similar to the experimental results of $\kappa_T$ (fig. 1b), where we also resolve a shoulder towards the liquid condensed phase and a steep change in $\kappa_T$ when leaving the liquid expanded phase. The enormous predictive power of the Eqs. (7)-(10) can be demonstrated by calculating the thermal expansion from the compressibility using $\Delta \kappa_T = -\frac{\Delta \alpha_\pi}{B_\pi}$. When taking for example $\kappa_T^{max}$ at 13°C (285K) and 23mN/m (fig. 1b), we find $\alpha_\pi \approx 0,4 \, K^{-1}$, which is close to $\alpha_\pi^{max} = 0,35 \, K^{-1}$ (at 13.6°C) extracted from the expansion measurement directly (3b). This quantitative and qualitative agreement indicates that the extrapolation from $\Delta A$ to area fluctuations $\delta A$ used to derive equation 10 is well justified.

*Coupling between electrical and elastic properties during phase transition*

As $\Psi$ was measured for DMPC, the electrical capacity $C_T$ can be calculated according to $C_T = \frac{dq}{d\Psi} = -\frac{\partial A}{\partial \Psi}\bigg|_{q,T} \frac{\partial \pi}{\partial \Psi}\bigg|_{A,T}$ (where we have used the Maxwell relation: $-\frac{\partial A}{\partial \Psi}\bigg|_{q,T} = \frac{\partial q}{\partial \pi}\bigg|_{A,T}$ ). By using $\Psi_t(T)$ to determine $B_\Psi$ from a linear fit (Eq. 2b), $C_T$ can directly be compared to $\kappa_T$. Although the correlation is more pronounced in figures 3a and 3b, the coupling is clearly visible in Fig. 3c as well.

*Heat Capacity of lipid monolayers*

The isobaric heat capacity $c_\pi$ (compare to $c_p$ as measured by differential scanning calorimetry (DSC) for small unilamelar lipid vesicles (SUVs)) of lipid monolayer is not directly experimentally accessible [12]. However using the relation shown in equation 10 we are now able to calculate $c_\pi(T)$ from $\alpha_\pi$.

The result plotted in Figure 3b contains a pronounced maximum, which qualitatively resembles the heat capacity profile of a lipid vesicle suspension. Remarkably, considering the calculated monolayer heat capacity profile at 32mN/m both the half width at half height (HWHH) as well as the absolute value of the maximum in $\Delta c_\pi(T)$ agree within 20% with the heat capacity profile of lipid SUVs of the same type [15]. Surprisingly, even the shoulder in the liquid ordered (gel) phase known from DSC measurements is found in our monolayer experiments. We believe that further theoretical investigations on this observation might give new insight on topological and pure in plane effects on the heat capacity profile of lipid suspensions like SUVs, since lipid monolayers simply can not undergo morphological transitions as opposed to lipid vesicles [16, 17, 18, 19].

*The Role of the Critical Point.*

The general character of the linear relationship between $\alpha_\pi, \kappa_T, c_\pi$ and $C_T$ calls for a more profound understanding of the origin of this linear relationship. Obviously, the character of the first order transition (e.g. plateau width) becomes increasingly weaker as the temperature rises (Fig. 1a), a typical fingerprint for the existence of a critical point. The same holds when looking at the area jump in figure 3a: As the pressure is increased towards the critical point the abrupt change in area decreases. Indeed, it has been stated, that the lipid monolayer phase transition is only of weak first order in the vicinity of a second order transition [20, 21].

In summary one can say, that all experiments indicate the trend towards a critical point as the temperature is increased. Since the symmetry between liquid expanded and liquid condensed phase remains beyond the critical point, the first order transition must converge into a second order transition (tri-critical point) [8]. At this point the Clausius-Clayperon equation for second order phase transitions states [5]:

$$\frac{\partial \pi}{\partial T} = \frac{\Delta c_\pi}{AT\Delta \alpha_\pi} \qquad (11),$$

or

$$\frac{\partial \Psi}{\partial T} = \frac{\Delta c_\pi}{T\Delta C_T} \qquad (12).$$

The vicinity of a critical point therefore implies $\Delta c_\pi \propto \Delta \alpha_\pi \propto \Delta C_T$ (equations 11, 12) near the phase transition regime, once transition pressure $\pi_t$ and potential $\Psi_t$ increase nearly linear with temperature *T*.

**Conclusions**

Our results reveal a thermal-mechano-electrical coupling in lipid monolayers. Using Einstein's theory of fluctuations in line with the experimental results, a linear relationship between electrical capacity, heat capacity, compressibility, and the thermal expansion coefficient has been found. These results enable us for the first time to calculate the heat capacity of lipid monolayers which has been shown to be within reasonable range with the heat capacity known from lipid vesicle suspensions. The relationship established in this work carries an enormous predictive power allowing to calculate heat and electrical properties from mechanical measurements. In future, it will be of enormous interest whether the same couplings hold for more complex or biological systems.

**Acknowledgement**
M.F.S. likes to thank Dr. K. Kaufmann for stimulating discussions and lectures concerning the thermodynamic foundation of soft interfaces. Furthermore, we would like to acknowledge very helpful remarks from Prof. Erich Sackmann. Financial support is acknowledged from the the German government through the Cluster of Excellence "NIM" and the DFG (SPP 1313). M.F.S., D.S. likes to thank the Fond der Chemischen Industrie as well as the Elitenetzwerk Bayern (CompInt).

**Figure 1**

a), b) DMPC pressure-area isotherms measured for different temperatures. For the sake of clarity only 4 isotherms are shown. From the raw data of the upper panel, the isothermal compressibility is calculated, showing a pronounced maximum during the phase transition regime (lower panel).
c) DMPC potential-area isotherms measured for different temperatures. The region around phase transition is plotted in a separate inset for every temperature, demonstrating the disappearance of the plateau with increasing temperature.

**Figure 2**

a) Plot of the transition pressure $\pi_t$ of the phospholipids DOTAP, DMPC, DMPG and a multicomponent system consisting of 40mol% DMPC, 40mol% DMPG and 20mol% DPPC as a function of temperature *T*. A linear relationship $\pi_t(T)$ is clearly conserved for all systems.
b) Transition Potentials $\Psi_t$ of DMPC as a function of temperature. Clearly, the surface potential $\Psi_t(T)$ increases linearly with temperature in transition regime.

**Figure 3**
a) By keeping the lateral pressure constant and varying the temperature, the area per molecule can be measured as a function of temperature. During the phase transition, the slope of the graph significantly increases.
b) From 3a the isobaric expansion coefficient $\alpha_\pi$ can be calculated. Similar to the heat capacity profiles, calculated using Eq. 11, it displays a maximum in the phase transition regime. The graphs display qualitatively and quantitatively (heat of transition) very similar features as the heat capacity obtained by DSC for giant unilamellar vesicles.
c) Our relations (Eq. 10) allow to directly compare the electrical capacity $C_T$ with the mechanical compressibility $\kappa_T$ at 8 °C. The compressibility as calculated from the electrical capacity $C_T$ using Eqs. 7, 8 and 9 is plotted in the same graph as the actual measured (using pressure-area isotherms) compressibility. Both curves show the typical fingerprints of a phase transition.

(a)

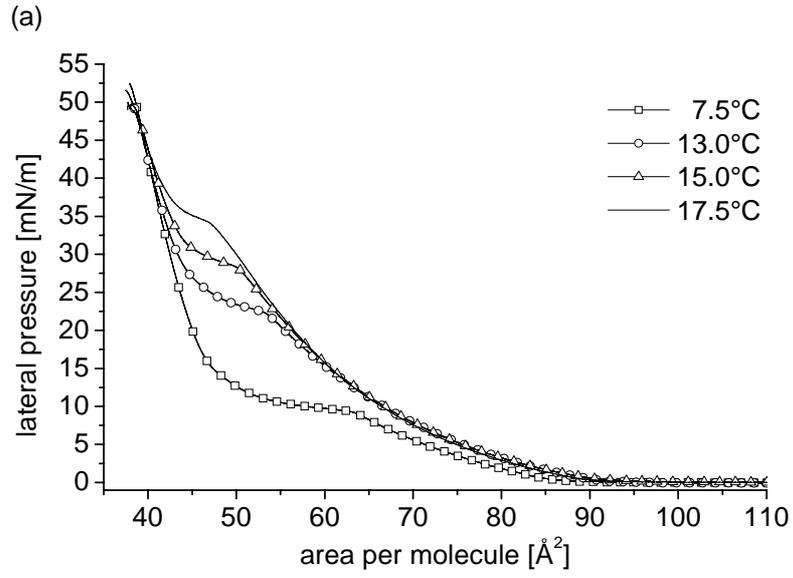

Figure 1

(b)

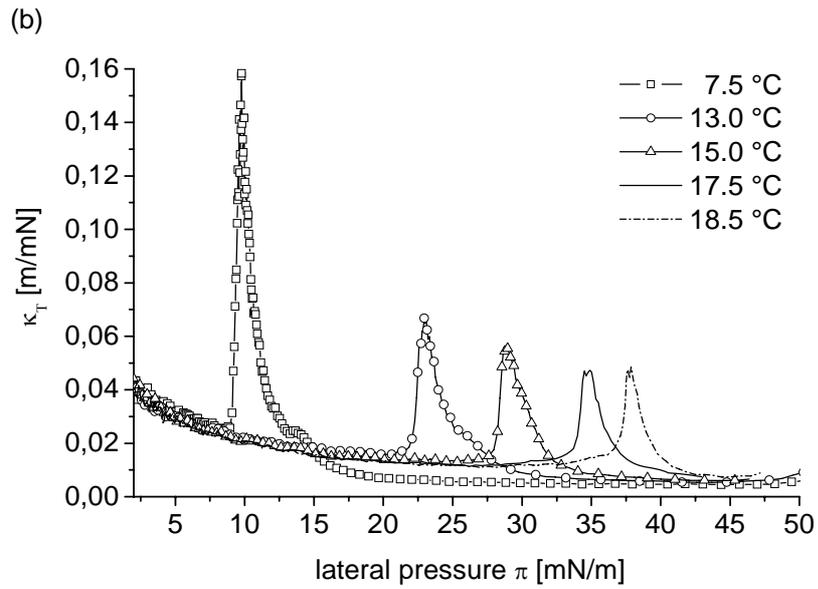

Figure 1

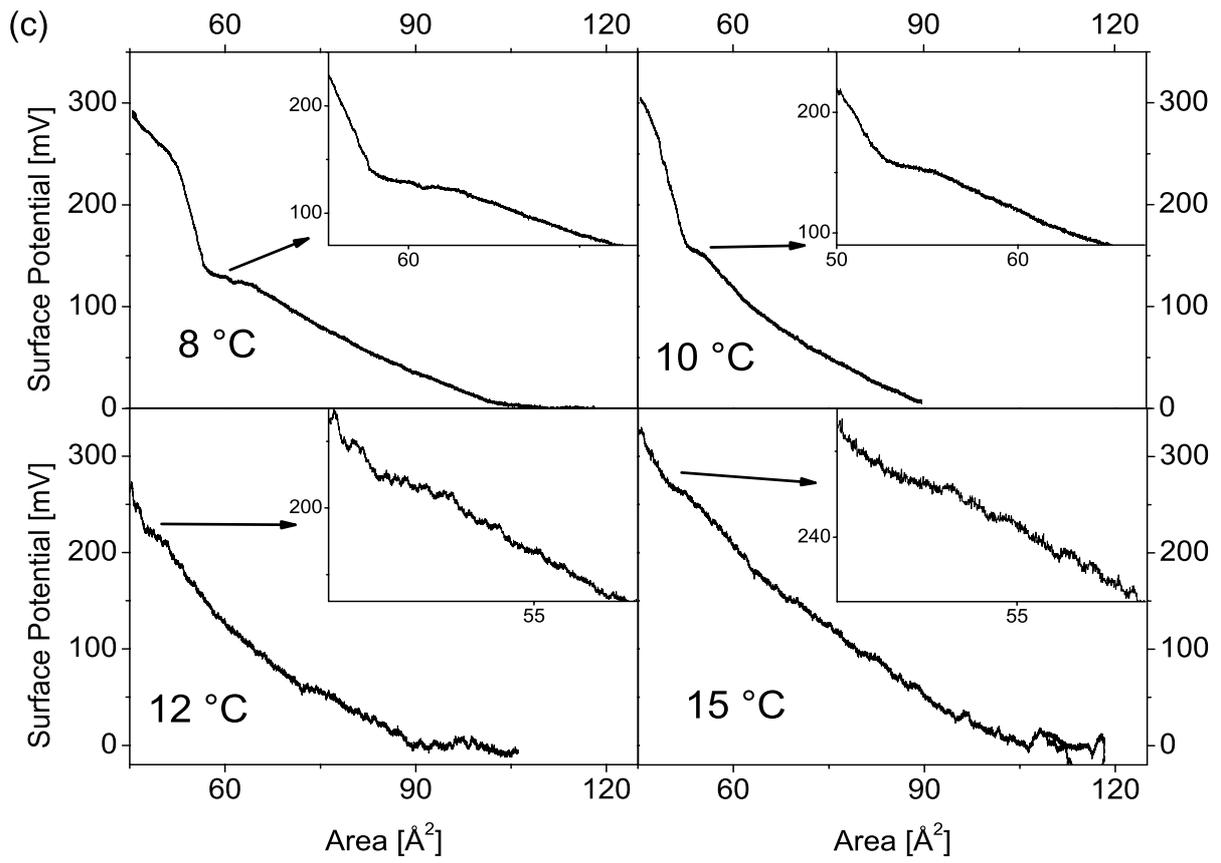

Figure 1

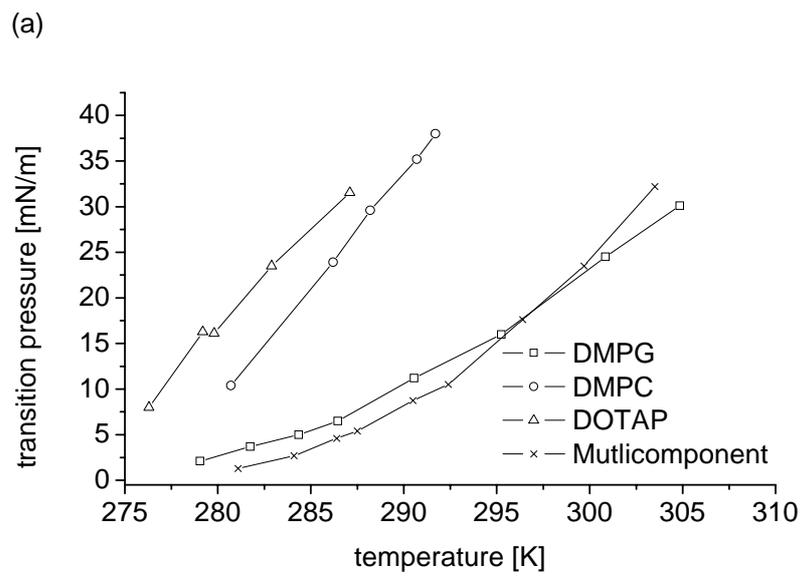

Figure 2

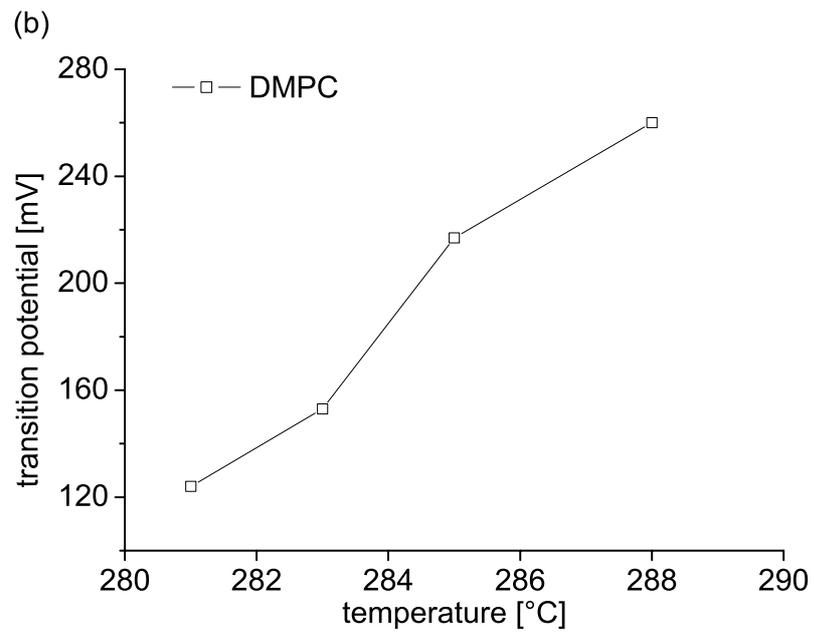

Figure 2

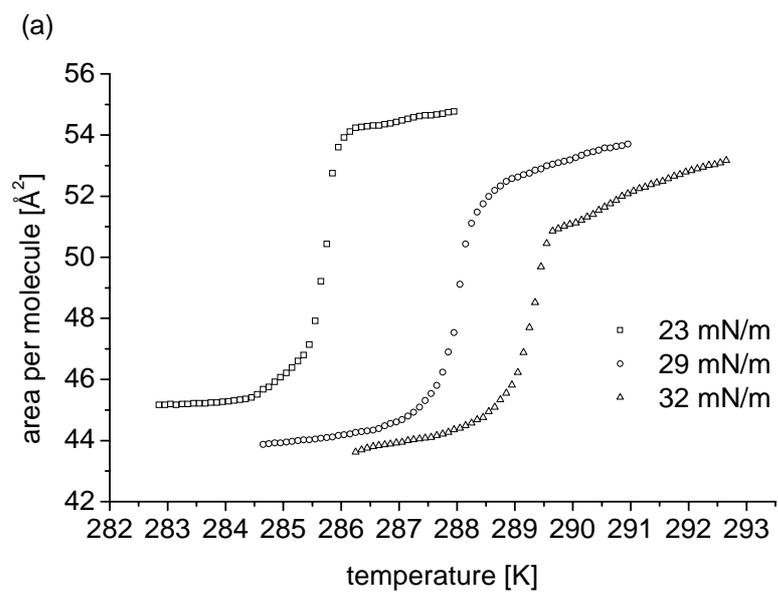

Figure 3

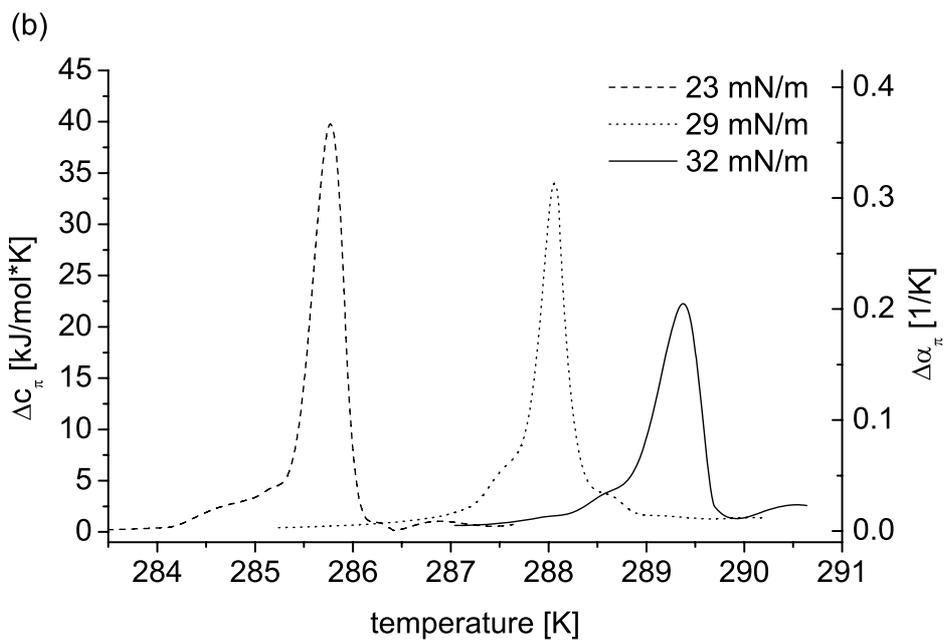

Figure 3

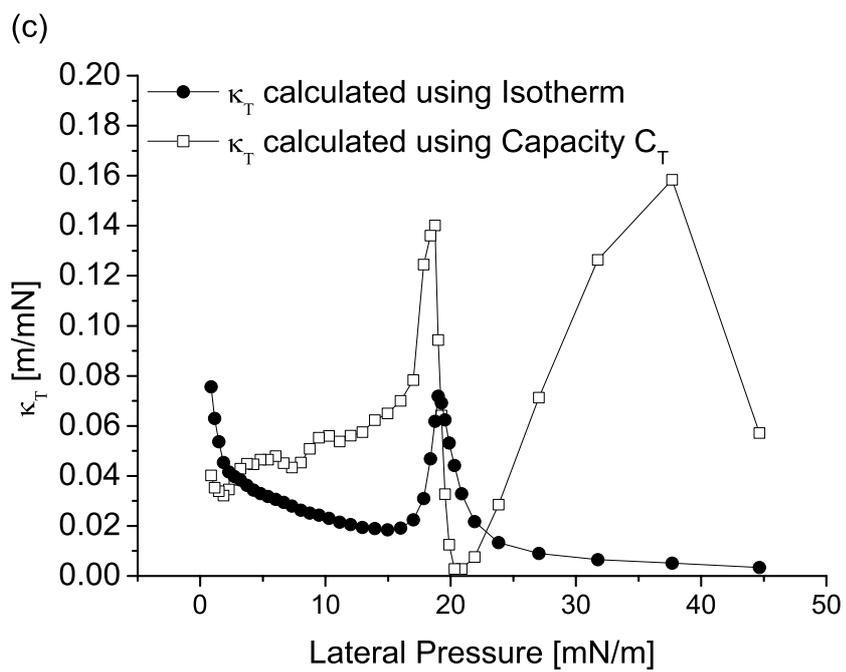

Figure 3